\documentclass[aps,superscriptaddress,floats,showpacs,twocolumn]{revtex4}
\usepackage{bm}
\usepackage{graphicx}
\usepackage{subfigure}

\begin{document}

\newcommand{\kb}{k_{\mathrm{B}}}
\newcommand{\etak}{\eta_{\mathrm{K}}}
\newcommand{\Rey}{\mathrm{Re}}
\newcommand{\Wi}{\mathrm{Wi}}
\newcommand{\St}{\mathrm{St}}
\newcommand{\taust}{\tau_\mathrm{st}}
\newcommand{\rhop}{\rho_{\mathrm{p}}}
\newcommand{\rhof}{\rho_{\mathrm{f}}}
\newcommand{\epsin}{\varepsilon_{\mathrm{in}}}
\newcommand{\epsp}{\varepsilon_{\mathrm{p}}}
\newcommand{\mb}{m_{\mathrm{b}}}
\newcommand{\np}{N_{\mathrm{p}}}
\newcommand{\la}{\langle}
\newcommand{\ra}{\rangle}

\title{Lagrangian tracer dynamics in a closed cylindrical turbulent convection cell }
\author{Mohammad S. Emran}
\affiliation{Institut f\"ur Thermo- und Fluiddynamik, Technische Universit\"at Ilmenau,
             Postfach 100565, D-98684 Ilmenau, Germany}
\author{J\"org Schumacher\footnote{Corresponding author: joerg.schumacher@tu-ilmenau.de}}
\affiliation{Institut f\"ur Thermo- und Fluiddynamik, Technische Universit\"at Ilmenau,
             Postfach 100565, D-98684 Ilmenau, Germany}
\date{\today}

\begin{abstract}
Turbulent Rayleigh-B\'{e}nard convection in a closed cylindrical cell is studied in 
the Lagrangian frame of reference with the help of three-dimensional direct numerical 
simulations. 
The aspect ratio of the cell $\Gamma$ is varied between 1 and 12, and the Rayleigh 
number $Ra$ between $10^7$ and $10^9$. The Prandtl number $Pr$ is fixed at 
0.7. It is found  that both the pair dispersion of the Lagrangian  tracer particles and 
the statistics of the acceleration components measured along the particle trajectories 
depend on the aspect ratio for a fixed Rayleigh number for the parameter range covered 
in our studies. This suggests that large-scale  circulations present in the convection cell 
affect the Lagrangian dynamics. Our findings are in qualitative agreement with existing 
Lagrangian laboratory experiments on turbulent convection.    
\noindent 
\end{abstract}

\maketitle

\section{Introduction}
The motion of a fluid can be described either in the Eulerian or Lagrangian frame of reference. 
In case of fluid turbulence, the Lagrangian point of view brought interesting and new 
insights on the small-scale structure and statistics (see e.g. Yeung \cite{Yeung2002}, Toschi and 
Bodenschatz \cite{Toschi2009} for comprehensive reviews). These studies included intermittency 
of temporal velocity increments and accelerations \cite{Lap1,Mo1,Vo1} or the geometry of particle tracks \cite{Bra1}. 
Furthermore,  two-particle and multi-particle dispersion have been analysed and compared with 
classical predictions by Batchelor \cite{Batchelor1950} and Richardson \cite{Richardson1926}. 
 
Almost all of the experimental and numerical investigations in turbulent thermal convection have 
been conducted in the Eulerian frame. This includes studies on the turbulent heat transfer and 
large-scale circulations \cite{Ahlers2009,dpa} as well as on small-scale structures and dynamics 
of thermal plumes \cite{Lohse2010,Cast1,Emran2008}. Recently, first Lagrangian laboratory experiments on convection were conducted in a closed cylindrical vessel \cite{Gasteuil2007} and   numerical simulations were carried out in Cartesian slabs with periodic side walls and free-slip boundary conditions at the top and 
bottom \cite{Schumacher2008,Schumacher2009}. The focus of these studies was on the local variations 
of the heat transfer which can be measured along the trajectories of the Lagrangian tracers. In the 
numerical studies, these results could be connected to the local tracer accelerations. In the 
laterally infinitely extended layer (i.e. the configuration with periodic side walls), slightly less intermittent 
vertical accelerations $a_z$ were detected in comparison to the lateral ones, $a_x$ and $a_y$. 
This was manifested in the sparser tails of the probability density function (PDF) of $a_z$.   
   
In the present work, we want to take these numerical Lagrangian studies to the next level. We report 
investigations in a closed cylindrical cell with no-slip boundary conditions at all walls. This is the 
characteristic setup for almost all laboratory experiments.
It will allow for a closer comparison with the work by Gasteuil {\it et al.} \cite{Gasteuil2007}. 
In addition, we want to explore systematically how the finite size of the cell affects the Lagrangian dynamics, 
such as the long-time behaviour of the two-particle dispersion. The latter point is also 
motivated by recent 
Eulerian studies, in which a  systematic dependence of the turbulent heat transfer on the 
aspect ratio $\Gamma$ is found for small and moderate values of $\Gamma$ \cite{Bailon2009}. The dependence is in line with morphological changes in the large-scale circulation (LSC), e.g. from one-roll 
to multi-roll patterns, in the convection cell. 
Our boundary conditions will differ from those in Refs.~\cite{Schumacher2008,Schumacher2009}. 
This difference might also affect the statistics of the acceleration 
components. It is thus another open question if a different velocity boundary layer structure has an impact on the  intermittency of lateral and vertical acceleration components. 
     
The outline of the paper is as follows. In section II we present the equations of motion, the numerical
model and the tracer advection scheme for the non-uniform cylindrical mesh. Section III is 
focussed on the particle dynamics and local heat transfer followed by studies of the pair dispersion
in Section IV and the aspect ratio dependence of the acceleration statistics in Section V. 
Finally, we summarize  our findings and give a brief outlook.  

\section{Equations of motion and numerical model}
\subsection{Boussinesq equations}
We solve the Boussinesq equations for Rayleigh-B\'{e}nard convection numerically with 
a second-order finite difference scheme in cylindrical coordinates $(\phi,r,z)$ \cite{Verzicco1996,Verzicco2003}.  
The equations are given by 
\begin{eqnarray}
\label{nseq}
\frac{\partial{\bm u}}{\partial t}+({\bm u}\cdot{\bm\nabla}){\bm u}
&=&-{\bm \nabla} p+\nu {\bm \nabla}^2{\bm u}+\alpha g T {\bm e}_z\,,\\
\label{ceq}
{\bm \nabla}\cdot{\bm u}&=&0\,,\\
\frac{\partial T}{\partial t}+({\bm u}\cdot{\bm \nabla}) T
&=&\kappa {\bm \nabla}^2 T\,,
\label{pseq}
\end{eqnarray}
where $p({\bm x},t)$ is the (kinematic) pressure, ${\bm u}({\bm x},t)$ the velocity field, 
$\alpha$ the thermal expansion coefficient and $g$ the gravity acceleration.  The 
velocity field has a no-slip boundary condition at all walls. The temperature field is 
isothermal at the top and bottom plates and adiabatic at the side wall.  Following Refs. 
\cite{Emran2008,Bailon2009}, we have chosen the DNS grid such that the {\em global} maximum 
of the geometric mean of the grid spacings in $\phi,r$ and $z$, i.e. 
\begin{equation}
\Delta=\max_{\phi,r,z}[\sqrt[3]{r \Delta_{\phi}\Delta_r\Delta_z}]\,,
\label{maxgrid}
\end{equation}  
satisfies the resolution threshold by Gr\"otzbach \cite{Grotzbach1983}, namely 
$\pi\eta/\Delta>1$ where $\eta$ is the Kolmogorov dissipation length which is calculated
via the mean energy dissipation rate 
$\langle\epsilon\rangle=\frac{\nu^3}{H^4}(Nu-1)Ra Pr^{-2}$. The details are provided in Tab.1. 

The resolutions of the runs are exactly the same as in \cite{Bailon2009}. Note that the turbulence
is strongly inhomogeneous with respect to the vertical direction $z$. This results in amplitudes of
the energy dissipation rate which are by orders of magnitude larger close to the plates than in the bulk. 
A local Kolmogorov scale is therefore much larger in the bulk as in the boundary 
layers. It motivated us in \cite{Bailon2009} to relate the grid spacing to 
$\eta(z)=\nu^{3/4}/\langle\epsilon(z)\rangle_{A,t}$ rather than $\eta$.

The dimensionless parameters and their range of values in our numerical study are:
the Rayleigh number $Ra=\alpha g (T_{bottom}-T_{top})H^3/(\nu\kappa)$ between $10^7$
and $10^9$, the Prandtl number $Pr=\nu/\kappa=0.7$ and the aspect ratios  $\Gamma=D/H$
between 1 and 12. $D$ is the cell diameter and $H$ is the height of the cell.
\begin{table}
\renewcommand{\arraystretch}{1.2}  
\begin{center}
\begin{small}
\begin{tabular}{cccccc}
\hline \hline
$Run$ & $N_{\phi}\times N_r\times N_z$ & $Ra$ & $\Gamma$ & $\pi\eta/\Delta$ &  $N_{BL}$ \\
\hline
1 & $193\times97\times128$   & $10^7$  & 1 & 2.26 & 14\\
2 & $257\times165\times128$  & $10^7$ & 3 & 1.43 & 14\\
3 & $361\times257\times128$  & $10^7$ & 5 & 1.32 & 14\\
4 & $401\times311\times128$  & $10^7$ & 8 & 1.06 &  14\\
5 & $601\times401\times128$  & $10^7$ & 12 & 1.01& 14\\
6 & $271\times151\times256$  & $10^8$ & 1 & 1.73 & 20\\
7 & $361\times181\times310$  & $10^9$ & 1 & 1.03 & 17\\
\hline \hline
\end{tabular}
\caption{Simulation parameters for various $Ra$ and $\Gamma$ with a fixed $Pr=0.7$. 
The grid resolution, the Rayleigh number $Ra$, the aspect ratio $\Gamma$, the ratio 
$\pi\eta/\Delta$, and $N_{BL}$, the number of horizontal grid planes inside the thermal 
boundary layer thickness, are given. $N_{BL}$ satisfies the criteria discussed in 
\cite{Stevens2010} very well.  \label{tab0}} 
\end{small}
\end{center}
\end{table}

\subsection{Numerical integration scheme}
We conduct direct numerical simulations (DNS) of the Boussinesq equations. The 
equations are solved in a cylindrical coordinate frame. The spatial discretization is 
performed on a staggered grid and solved by a second-order finite difference scheme 
\cite{Verzicco1996,Verzicco2003}. The pressure field $p$  is determined by a two-dimensional 
Poisson solver \cite{fish} after applying one-dimensional fast Fourier transformations 
in the azimuthal  direction. The time advancement is done by a third-order Runge-Kutta 
scheme. The grid spacings are non-equidistant in the radial and axial directions. In the 
vertical direction, they correspond to Tschebycheff collocation points. Some parameters 
and the corresponding grid resolutions are given in Table \ref{tab0}.

\subsection{Tracer advection}
The equations of Lagrangian tracer motion are given by  
\begin{equation}
\frac{d\bm x_p (t)}{dt} = \bm u(\bm x_p,t)\,.
\label{eq:tra}
\end{equation}
Here $\bm u = (u_{\phi}(\phi_p), u_r(r_p), u_z(z_p))^T$ is the velocity of the particle at position 
$\bm x_p = (\phi_p(t), r_p(t), z_p(t))^T$ with the symbols 
$\phi$, $r$ and $z$ corresponding to the azimuthal, radial and axial directions, respectively.
\begin{figure}
\begin{center}
\includegraphics[angle=0,scale=0.6,draft=false]{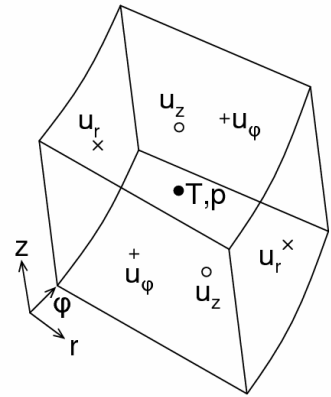}
\caption{Sketch of a 3D grid cell in the cylindrical coordinates, in which the positions of 
the velocity components, the temperature and the pressure are shown on the staggered grid.} 
\label{fig:Lgrid1}
\end{center}
\end{figure}
\begin{figure}
\begin{center}
\includegraphics[angle=0,scale=0.5,draft=false]{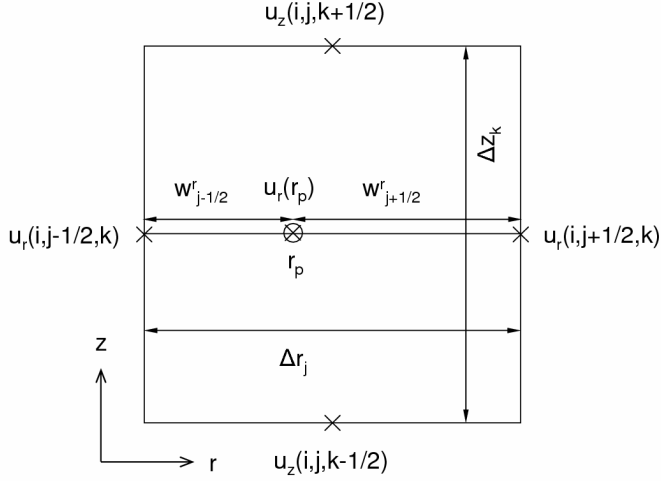}
\caption{Position of the velocity components on a 2D staggered grid in $r-z$ plane, 
which is the mid section of a typical 3D cell as shown in the top panel of the figure. 
The indices $i$, $j$ and $k$ correspond to the azimuthal, radial and axial directions, 
respectively. The particle velocity $u_r(r_p)$ at position $r_p$ is obtained by the linear 
interpolation of the Eulerian velocities $u_r(i,j-\frac{1}{2},k)$ and $u_r(i,j+\frac{1}{2},k)$ with the 
widths $w^r_{j-\frac{1}{2}}$ and $w^r_{j+\frac{1}{2}}$ as given in (\ref{eq:lz}).} 
\label{fig:Lgrid2}
\end{center}
\end{figure}
As discussed before, the Eulerian fields are well resolved on a staggered and non-uniform 
grid in the simulations. The velocity components, as shown in Fig.~\ref{fig:Lgrid1},  
are then given on particular faces of the grid cells only, while the Eulerian temperature field is 
always centered in the grid cell. All these fields are required at the Lagrangian particle position to
advance (\ref{eq:tra}) and study local heat transfer. We therefore applied an interpolation scheme for the velocity components which makes direct use of the staggered setup. The interpolation is then of 
second order and the maximal error can be estimated to be of the order of ${\cal O}(\Delta^2)$ 
(see (\ref{maxgrid})). Given the Eulerian velocities at the mid points of the edges of the mesh 
cell  (see Fig.~\ref{fig:Lgrid2}), we calculate the corresponding Lagrangian 
components  at an arbitrary position inside a cell as
\begin{eqnarray} \label{eq:lx}
u_r(r_p) &=& \frac{w^r_{j-\frac{1}{2}} u_r(i,j+\frac{1}{2},k) + w^r_{j+\frac{1}{2}} u_r(i,j-\frac{1}{2},k)}{\Delta_r(i,j,k)} \,,\nonumber \\
u_{\phi}(\phi_p) &=& \frac{w^{\phi}_{i-\frac{1}{2}} u_{\phi}(i+\frac{1}{2},j,k) + w^{\phi}_{i+\frac{1}{2}} u_{\phi}(i-\frac{1}{2},j,k)}{\Delta_{\phi}(i,j,k)} \,,\nonumber \\
u_z(z_p) &=& \frac{w^z_{k-\frac{1}{2}} u_z(i,j,k+\frac{1}{2}) + w^z_{k+\frac{1}{2}} u_z(i,j,k-\frac{1}{2})}{\Delta_z(i,j,k)}\,.
\label{eq:lz}
\end{eqnarray}
Indices $i, j, k$ denote the grid positions with respect to $\phi, r, z$, which run from 1 to 
$N_{\phi}$, $N_r$  and $N_z$, respectively. 

In order to progress further, we rewrite the three ordinary differential equations
(we provide here details of the equation in $r$-direction only) \cite{Schonfeld1995} as
\begin{equation}
\frac{d r_p}{dt} = u_r(r_p)  = \frac{r_p -r_L}{r_R-r_L} u_{r,R} + \left( 1- \frac{r_p -r_L}{r_R-r_L}
\right)u_{r,L} \,,
\label{eq:xp}
\end{equation}
where $u_{r,R}$ and $u_{r,L}$ are the Eulerian velocities at face positions $r_R$ and $r_L$, 
respectively and $r_p$ is the particle's position in between. The subscripts $L$ and $R$ 
refer to the left and right faces of a 3D-cell (see Fig.~\ref{fig:Lgrid1}), respectively. 
Rearranging (\ref{eq:xp}) yields \cite{Schonfeld1995}
\begin{equation}
\frac{d r_p}{dt} - r_p \frac{\Delta u_r}{\Delta r} =u_{r,L} - r_L\frac{\Delta u_r}{\Delta r} \,,
\label{eq:xpn}
\end{equation}
with $\Delta u_r=u_{r,R}-u_{r,L}$ and $\Delta r=r_R-r_L$. Multiplying both sides of 
(\ref{eq:xpn}) with the integrating factor $\exp(-\frac{\Delta u_r}{\Delta r} t)$ gives
\begin{equation}
\frac{d }{dt}\left( r_p\,e^{-\frac{\Delta u_r}{\Delta r} t} \right)  =\left(  u_{r,L} - r_L \frac{\Delta u_r}{\Delta r} \right) 
e^{-\frac{\Delta u_r}{\Delta r} t} \,.
\label{eq:xpi}
\end{equation}
Now integrating both sides of (\ref{eq:xpi}) with respect to $t$ gives the $r$-position 
of the particle as
\begin{equation}
r_p(t)  =  r_L - u_{r,L} \frac{\Delta r}{\Delta u_r} + C\, e^{\frac{\Delta u_r}{\Delta r} t} \,.
\label{eq:xt}
\end{equation}
At $t=t_0$
\begin{equation}
r_p(t_0)  =  r_L - u_{r,L} \frac{\Delta r}{\Delta u_r} + C\, e^{\frac{\Delta u_r}{\Delta r} t_0} \,,
\label{eq:x_0}
\end{equation}
and thus
\begin{equation}
\delta r =r_p(t) - r_p(t_0)  = C \left(  e^{\frac{\Delta u_r}{\Delta r} t} - e^{\frac{\Delta u_r}
{\Delta r} t_0}\right)  \,.
\label{eq:delr}
\end{equation}
From Eq.~(\ref{eq:x_0}), with $r_p(t_0)=r_0$, and Eq. (\ref{eq:xpn}) at $t=t_0$ the constant $C$ follows
to
\begin{equation}
C =   u_{r,0} \frac{\Delta r}{\Delta u_r} e^{-\frac{\Delta u_r}{\Delta r} t_0} \,.
\label{eq:rc}
\end{equation}
and thus the radial increment for $\delta t= t-t_0$ is
\begin{equation}
\delta r  = u_{r,0} \frac{\Delta r}{\Delta u_r} \left( e^{\frac{\Delta u_r}{\Delta r} \delta t} -1 \right) 
\label{eq:del_radial}
\end{equation}
Expanding the exponential term, this scheme is first order in time for the leading contribution. 
Similarly, for the azimuthal and axial directions, the distances are given by
\begin{eqnarray}
\delta \phi  &=& \frac{u_{\phi,0}}{r_0} \frac{\Delta \phi}{\Delta u_{\phi} } \left( e^{\frac{\Delta u_{\phi}}
{\Delta \phi} \delta t} -1 \right) \,, \\
\delta z  &=& u_{z,0} \frac{\Delta z}{\Delta u_z} \left( e^{\frac{\Delta u_z}{\Delta z} \delta t} -1 \right)   \,.
\label{eq:del_axial}
\end{eqnarray}
The equations of the particle trajectories in the Lagrangian frame are
\begin{eqnarray}
r_p(t) = r_p(t_0) +  \delta r  \,,\\
\phi_p(t) = \phi_p(t_0) +  \delta \phi  \,,\\
z_p(t) = z_p(t_0) +  \delta z  \,.
\label{eq:traj}
\end{eqnarray}

The temperature field $T$ is stored at the cell center of the staggered grid as shown in
Fig.~\ref{fig:Lgrid1}. At first, we interpolate $T$ at the vertices of the Eulerian grid and then 
apply trilinear interpolation to calculate the Lagrangian temperature at a known particle 
position.

\section{Tracer dynamics in the cell}
\subsection{Tracer trajectories and large-scale circulation}
For the Lagrangian simulations, we have selected cylindrical cells which are given in
Tab.~\ref{tab0}  keeping the Prandtl number constant ($Pr=0.7$). We advect $N_p\sim 10^5$ 
tracer particles in the cell (see Tab. \ref{tab1} for more details). They are initially seeded in the whole
cell and integrated in time for more than $150$ time units $t/t_f$ where 
$t_f=H/U_f$ is the free fall time. Here, $U_f=\sqrt{g\alpha (T_{bottom}-T_{top}) H}$. The 
statistical analysis is thus conducted over $3000$ independent Lagrangian samples, which are separated by $\Delta t=0.05t_f$ which corresponds to $\Delta t=$ $0.25\tau_{\eta}$, $0.33\tau_{\eta}$ and
$0.43\tau_{\eta}$ for $Ra=10^7$, $10^8$ and $10^9$, respectively. 
\begin{table}
\renewcommand{\arraystretch}{1.2}  
\begin{center}
\begin{small}
\begin{tabular}{ccccccc}
\hline \hline
Run & $t/t_f$ & $Nu_E\pm\sigma$ & 
$Nu_L\pm\sigma$ & $N_p$  \\
\hline
1  &  150  & 16.73$\pm$0.08 & 16.77$\pm$2.95  & 104160\\
2  &  150  & 16.06$\pm$0.05 & 15.86$\pm$1.37  & 81024 \\
3  &  150  & 16.36$\pm$0.03 & 15.70$\pm$0.57  & 202860\\
4  &  150  & 17.44$\pm$0.02 & 16.00$\pm$0.54  & 208425 \\
5  &  150  & 17.49$\pm$0.03 & 16.06$\pm$0.54  & 272811 \\
6  &  150  & 32.21$\pm$0.32 & 31.75$\pm$4.29  & 112320\\
7  &  139  & 64.31$\pm$0.64 & 65.12$\pm$19.53& 179850\\
\hline \hline
\end{tabular}
\caption{The total 
integration time $t/t_f$, the Eulerian and Lagrangian values of the Nusselt number plus errorbars 
$Nu_E$ and $Nu_L$, and the total number of  tracers $N_p$ are listed. \label{tab1}} 
\end{small}
\end{center}
\end{table}

\begin{figure}[!ht]
\begin{center}
\includegraphics[angle=0,scale=0.32,draft=false]{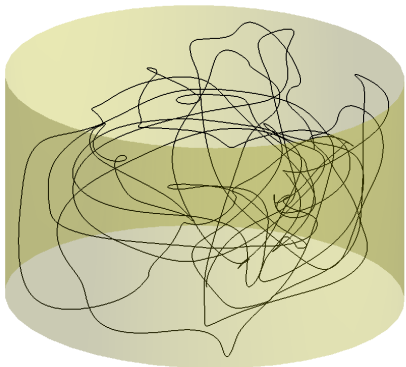}
\caption{(Color online) Three-dimensional trajectory of a Lagrangian tracer particle for the 
simulation at  $Ra=10^7$ in the cell with aspect ratio 
$\Gamma=1$ . The tracer track is displayed for 100 free-fall
time units $t_f$.} 
\label{fig:tracer_path}
\end{center}
\end{figure}
\begin{figure}[!ht]
\begin{center}
\includegraphics[angle=0,scale=0.6,draft=false]{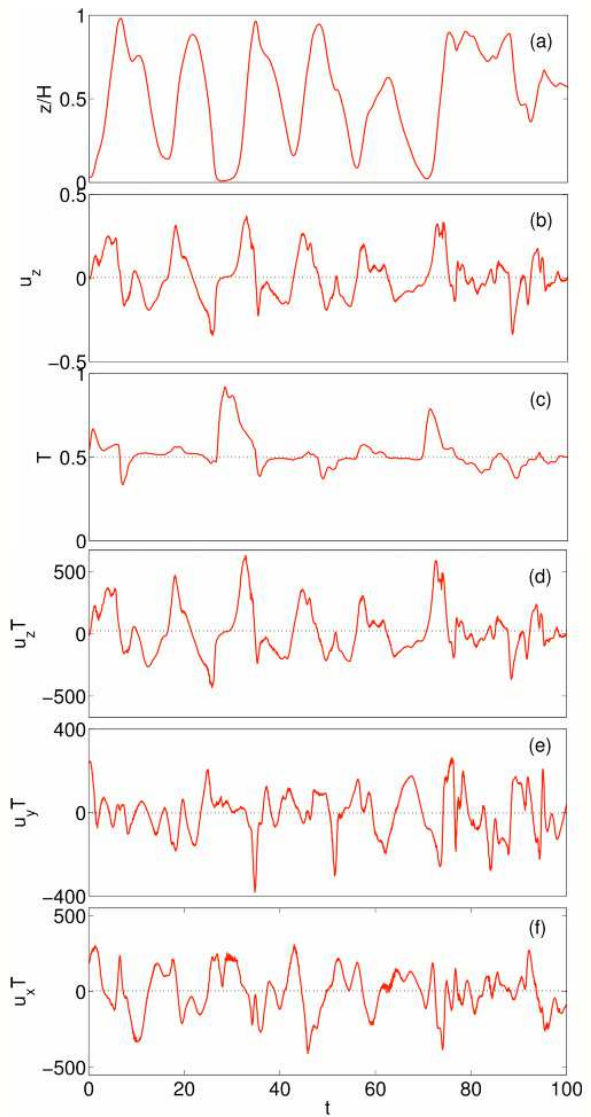}
\caption{(Color online) Time series of various quantities along the trajectory 
of a single Lagrangian tracer particle: (a)  the vertical position $z(t)$ 
normalized by the cell height $H$, (b) the vertical velocity $u_z(t)$, (c) 
the temperature $T(t)$, products of velocity components and temperature 
(d) $u_zT$, (e) $u_yT$ and (f) $u_xT$. The dotted lines in some of the panels 
show the corresponding  value of the time average which enters the Lagrangian
definition of the Nusselt number as given in Eq.~(\ref{eq:NuLa}). Data are 
taken along the tracer track which is shown in Fig. \ref{fig:tracer_path}}
\label{fig:tracer_t}
\end{center}
\end{figure}

We visualize the time-evolution of the trajectory of a Lagrangian particle in Fig.~\ref{fig:tracer_path} for a 
simulation at $Ra=10^7$ in the Rayleigh-B\'enard cell with aspect ratio $\Gamma=1$. It seems 
that this particle travels preferentially along the large-scale circulation (LSC). Fig.~\ref{fig:tracer_t} 
plots the time series of the vertical position and various quantities along the same particle path as shown in 
Fig.~\ref{fig:tracer_path}. The results are qualitatively in agreement with the Eulerian measurements
\cite{Shang2004} and the Lagrangian measurements \cite{Gasteuil2007}. Sharp spikes of the 
temperature signal appear due to rising and falling plumes close to the hot and cold plates of the 
cell. They are correlated with the maxima and minima in the vertical position. The vertical velocity 
$u_z(t)$ has zero mean. The convective heat fluxes are shown in Fig.~\ref{fig:tracer_t}(d)-(f). The signals are highly irregular with sharp intermittent peaks. The factor $\sqrt{Ra Pr}$ is multiplied to all flux terms 
$u_xT$, $u_yT$ and $u_zT$ in order to get the values of the dimensionless local convective heat flux. 
Similar to Gasteuil {\it et al.} \cite{Gasteuil2007} we observe strong fluctuations about the mean value
which are $4.8$, $-1.7$ and $23.4$ for $u_xT$, $u_yT$ and $u_zT$, respectively. It demonstrates 
that the vertical flux $u_zT$ is dominantly responsible for the net heat transfer in the cell.

The time series of the vertical tracer position $z(t)$ shows a rather regular oscillatory form which is 
caused by the mean wind in the cell. One can therefore extract a typical loop time from all the time 
series. Motivated by the analysis of Sreenivasan {\it et al.} \cite{Sreenivasan2002} for mean wind reversals,
we have therefore performed a statistics of the zero-crossings for $z_i(t)$ with $i=1.. N_p$.  The 
mean time interval between two subsequent zero crossings for each tracer track has been determined,
and a subsequent Lagrangian ensemble mean, $\langle\cdot\rangle_L$
has been taken. The latter, which we denote as  $\langle T_{zc}\rangle_L$, gives  a characteristic loop 
time of the tracers, $T_l=2\langle T_{zc}\rangle_L$. For a fixed Rayleigh number $Ra=10^7$,  $T_l= 19.4 t_f, 17.8 t_f$ and $19.0 t_f$ for $\Gamma=1, 3$ and 8, respectively.  

Translating into a convective time unit 
$t_c=H/\langle u^2\rangle_{V,t}^{1/2}$ (see Ref. \cite{Bailon2009} for a discussion)  gives 
$T_l\approx 3.7 t_c,  3.4 t_c$ and  3.6 $t_c$ for $\Gamma=1, 3$ and 8, respectively. For a fixed aspect ratio $\Gamma=1$, we find $T_l=19.6 t_f$ at $Ra=10^8$  and $T_l=20.6 t_f$ at $Ra=10^9$. We see that the characteristic loop time of the tracer is slightly varying but no clear trend when $\Gamma$ is varied and $Ra$ is fixed. It seems, therefore, to be independent of the presence of a single-roll or multi-roll LSC. In case of the fixed aspect ratio, a slight increase of $T_l$ with increasing $Ra$ is detected. We recall that the velocity fluctuations amplitude decreases with increasing $Ra$ as reported 
for example 
by Verzicco and Camussi \cite{Verzicco2003}. Our observation of a slightly longer loop time seems to be 
consistent with a reduced amplitude of velocity fluctuations with increasing $Ra$. On the basis of 
the present data, it is however difficult to draw a firm conclusion about the robustness of this trend.

 
\subsection{Heat transfer in the Lagrangian frame}
Similar to the Eulerian case, where the Nusselt number is given by 
\begin{equation}
Nu_E =1 + \frac{H}{\kappa\Delta T}\la u_zT \ra_{V,t}  \,,
\label{eq:NuEu}
\end{equation}
with $\la\cdot\ra_{V,t}$ being a volume and time average, the Nusselt number 
in the Lagrangian frame, $Nu_L$, is given by 
\cite{Schumacher2009}
\begin{equation}
 Nu_L =1 + \frac{H}{\kappa\Delta T}\la u_zT \ra_{L,t}=\la Nu_L(t)\ra_t  \,,
\label{eq:NuLa}
\end{equation}
\begin{figure}[!ht]
\begin{center}
\includegraphics[angle=0,scale=0.3,draft=false]{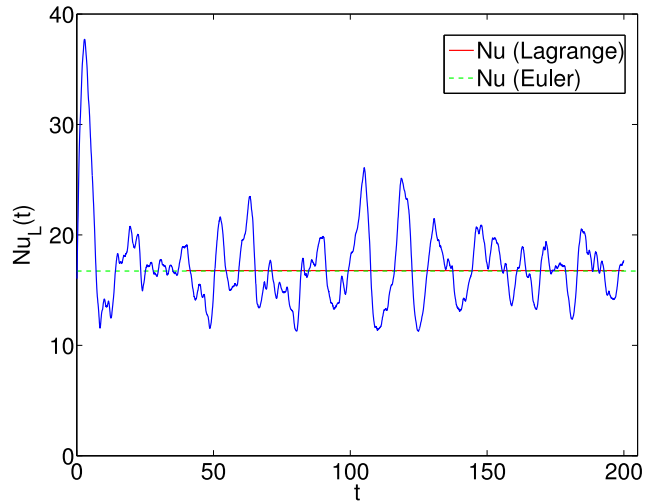}
\caption{(Color online) Graph of $Nu_L(t)$ which is averaged over 
$\sim 10^5$ tracer paths. The temporal mean value of $Nu_L=\la 
Nu_L(t)\ra_t$ (see Eq.~(\ref{eq:NuLa})) between time $t=40-200$ 
is $Nu_L=16.77$, which almost exactly matches with the corresponding 
Eulerian value $Nu_E=16.73$ as listed in Table~\ref{tab1}.} 
\label{fig:NuLag}
\end{center}
\end{figure}
where the symbol $\la\cdot\ra_{L,t}$ denotes an averaging over all trajectories and time. 
We compute the time series of the heat transfer in the Lagrangian 
frame and compare with its value in the Eulerian frame in Fig.~\ref{fig:NuLag} for a simulation 
at $Ra=10^7$ in the cell with aspect ratio $\Gamma=1$. The time averaging of  
$Nu_L(t)=1+H/(\kappa\Delta T) \la u_z(t)T(t)\ra_L$ 
curve from $t=40$ to 200 yields $Nu_L=16.77$, which almost exactly matches the corresponding 
value of the Eulerian simulation, which was $Nu_E=16.73$ as provided in Tab.~\ref{tab1}. Similar 
result is achieved for the simulation conducted at $Ra=10^8$ in the same cell, for which $Nu=32.21$ 
and $Nu_L=31.75$ with a deviation of 1.5\% from $Nu_E$. The results also validate our interpolation schemes and 
the time step sizes for the Lagrangian simulations, which are accurate enough to reproduce flow 
quantities of the corresponding Eulerian simulations in turbulent convection. In contrast, recent 
experiments by \cite{Gasteuil2007} obtained a value of $Nu_L$ almost twice as large as the  
Eulerian case. This discrepancy is due to the fact that their \cite{Gasteuil2007} mobile sensor traversed preferentially along the mean flow circulation path and hence the contribution from the rest of the volume was missing in their Nusselt number measurement.

At the beginning of the $Nu_L(t)$ curve in Fig.~\ref{fig:NuLag}, there is a large overshoot from mean, which generally appears when the majority of the particles follow the LSC path or the particles are not seeded uniformly in the flow. We also observe strong oscillations in
$Nu_L(t)$ for all the simulations listed in Tab.~\ref{tab1}. Strong fluctuations around the mean result in large standard deviation $\sigma$. The value of $\sigma$ decreases with increasing $\Gamma$. When the number of particles was increased (case: $N_p\approx 2.5\times10^5$, $Ra=10^7$ and $\Gamma=1$; the results are not shown here), the convergence of $Nu_L(t)$ was achieved within a shorter period of time. It can be concluded that the number of tracer particles should be sufficiently large and 
they should be uniformly seeded in the domain for faster statistical convergence of $Nu_L(t)$.
\begin{figure}
\begin{center}
\includegraphics[angle=0,scale=0.27,draft=false]{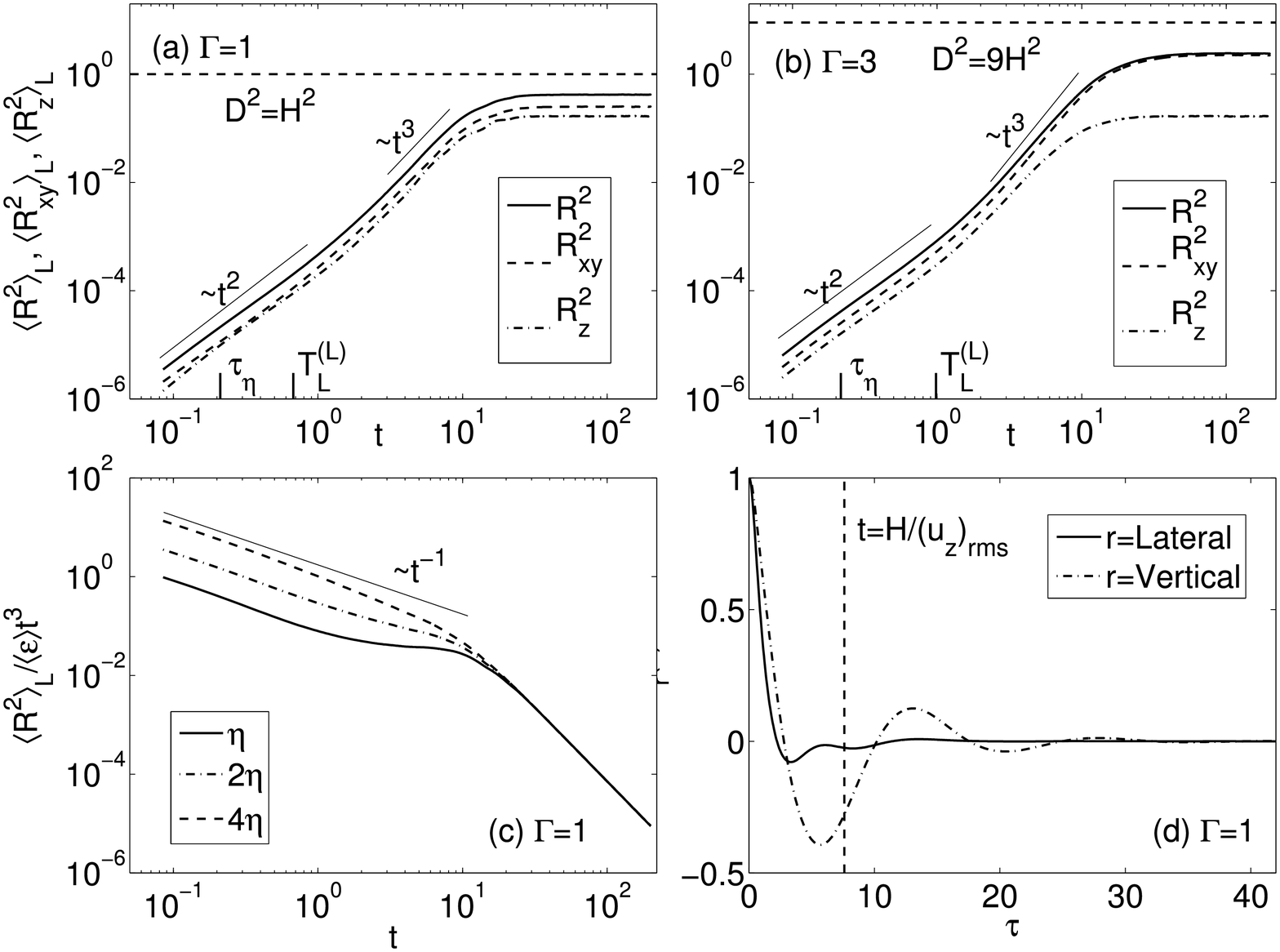}
\caption{Pair dispersion as in Eqns.~(\ref{eq:R2})--(\ref{eq:Rz}): 
(a) aspect ratio $\Gamma=1$, (b) $\Gamma=3$, (c) pair dispersion 
normalized by $\langle\epsilon\rangle t^3$ for three initial separations, 
$\eta$, $2\eta$ and $4\eta$, with $\eta$ the Kolmogorov length and 
(d) the autocorrelation functions of the lateral and vertical velocities 
for $\Gamma=1$ as given in Eqns.~(\ref{eq:Cxy}) and (\ref{eq:Cz}), 
respectively. The Kolmogorov time $\tau_{\eta}$ (see  Eq.~(\ref{eq:Tk})) 
and the lateral Lagrangian time $T^{(L)}_L$ (see  Eq.~(\ref{eq:Tla})) 
are marked on the time axis in (a) and (b). The horizontal dashed 
lines in (a) and (b) denote $D^2=\Gamma^2 H^2$. The vertical dashed 
line in (d) represents the characteristic time required by a particle to travel 
the cell of height $H$. The analysis is conducted at $Ra=10^7$ with  
$4.16\times10^8$ and  $3.24\times10^8$ statistical samples for $\Gamma=1$ and 
$\Gamma=3$, respectively.} 
\label{fig:tracer_R}
\end{center}
\end{figure}
\begin{figure}
\begin{center}
\includegraphics[angle=0,scale=0.35,draft=false]{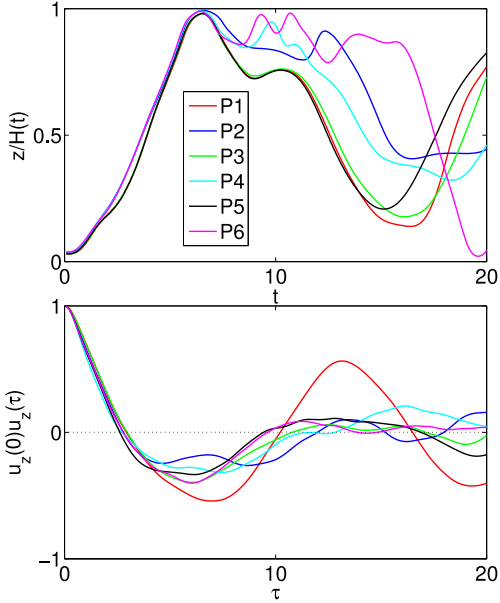}
\caption{(Color online) Relation between the vertical coordinate and the 
anti-correlation of vertical velocity. In the top panel, we show the 
$z(t)$-component of 6 particle trajectories, which were seeded 
close to the bottom plate. The corresponding product 
$u_z(0)u_z(\tau)$ along each trajectory which enters into the 
definition of the Lagrangian integral time is shown in the bottom panel.} 
\label{fig:tracer_corr}
\end{center}
\end{figure}
\begin{figure}
\begin{center}
\includegraphics[angle=0,scale=0.25,draft=false]{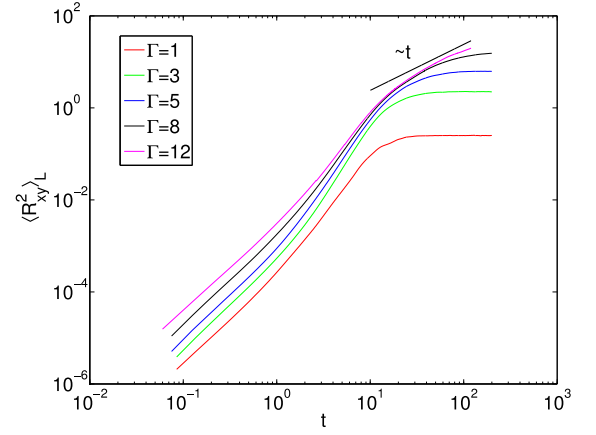}
\caption{(Color online) Time evolution of the lateral pair dispersion as a 
function of the aspect ratio $\Gamma$. Data are for $Ra=10^7$. The 
lowermost curve is for $\Gamma=1$ and the uppermost for 
$\Gamma=12$.} 
\label{fig:tracer_R_Gamma}
\end{center}
\end{figure}

\section{Lagrangian pair dispersion}
The pair dispersion measures the relative separation of a pair of tracer particles traversing along their 
trajectories and is defined as
\begin{equation}
R^2(t) =\la \left[\bm x_p(t) - \bm x_p(t_0)\right]^2 \ra_L \,.
\label{eq:R2}
\end{equation}
Since convective turbulence is inhomogeneous, we decompose the distance vector into two parts -- 
the lateral part $\bm R_{xy}(t)$ and the vertical part $\bm R_z(t)$. Their contributions to the total dispersion
$R^2(t)$ are given by 
\begin{eqnarray}
R^2_{xy}(t) &= & \la \left[x_p(t) -  x_p(t_0)\right]^2 + \left[y_p(t) - y_p(t_0)\right]^2 \ra_L \\
\label{eq:Rxy}
R^2_z(t) &= & \la \left[z_p(t) -  z_p(t_0)\right]^2 \ra_L \,.
\label{eq:Rz}
\end{eqnarray}
This decomposition is chosen in order to relate our results to the findings in \cite{Schumacher2008,Schumacher2009}.
We also compute the autocorrelation functions of the velocity components as 
\begin{eqnarray}\label{eq:Cxy}
C_{xy}(\tau) &= & \frac{\la \bm u_{xy}(t+\tau) \cdot \bm u_{xy}(t)\ra_{L,t}} {\la |\bm u_{xy}|^2\ra_{L,t}} \,, \\
C_z(\tau) &= & \frac{\la u_z(t+\tau) u_z(t)\ra_{L,t}}{ \la u_z^2 \ra_{L,t}} \,.
\label{eq:Cz}
\end{eqnarray}
Here $C_{xy}(\tau)$ is the autocorrelation coefficient for the lateral tracer velocities $\bm u_{xy} =
u_x \bm e_x + u_y \bm e_y$, with $\bm e_x$ and $\bm e_y$ are the unit vectors in $x$ and $y$ 
directions, respectively, $C_z(\tau)$ is for the vertical velocity $u_z$, and $\tau$ is the time lag. 
The vertical and  lateral Lagrangian  times, $T_L^z$ and  $T_L^{xy}$ are obtained by integrating 
$C_z(\tau)$ and $C_{xy}(\tau)$ as
\begin{eqnarray}
T_L^z&=&\int_0^{\infty} C_{z}(\tau)\,\mbox{d}\tau \,\\
T_L^{xy} &=&\int_0^{\infty} C_{xy}(\tau)\,\mbox{d}\tau \,,
\label{eq:Tla}
\end{eqnarray}
and the Kolmogorov time $\tau_{\eta}$ is given by
\begin{equation}
 \tau_{\eta} =\sqrt{\frac{\nu}{\la\epsilon\ra}} \,.
\label{eq:Tk}
\end{equation}
Figures~\ref{fig:tracer_R}(a) and (b) plot the total dispersion, $R^2$, as well as the lateral, $R^2_{xy}$, and 
vertical, $R^2_z$, contributions. Two aspect ratios are selected, namely $\Gamma=1$ and 3 for simulations 
at $Ra=10^7$. Initially, the quantities grow with $R^2\sim t^2$, which corresponds to the ballistic dispersion. 
After the ballistic growth, a transition to a regime with a growth law of $R^2\sim t^3$ occurs. 
Similar observations were made for fluid turbulence \cite{Boffetta2002} and for convective turbulence in an 
extended layer \cite{Schumacher2008,Schumacher2009}. The range is
however too short to conclude if this a Richardson-like regime \cite{Richardson1926} or not.
The horizontal dashed lines mark the square of the 
cell diameter, the limit which cannot be exceeded by pair separation. The finite size of the cell 
suppresses also an eventual Taylor diffusion regime with $R^2\sim t$ and  as a result the dispersion 
levels off, which is different from the Cartesian cells with periodic sidewalls \cite{Schumacher2008,Schumacher2009}. 
The Kolmogorov time $\tau_{\eta}$  and the lateral Lagrangian integral time $T_L^{xy}$ as a function of the aspect 
ratio are listed in Tab.~\ref{kolscale}. While the Kolmogorov time remains nearly unchanged with an increase of 
$\Gamma$, the lateral Lagrangian integral time increases by almost 100\% for an increase of $\Gamma$ from 1 to 5. 
This would be in line with a stronger correlation of the lateral velocity components which could arise in the multi-roll
LSC case.  
\begin{table}
\renewcommand{\arraystretch}{1.2}  
\begin{center}
\begin{tabular}{ccccc}
\hline \hline
$\Gamma$ & 1 & 3 & 5 & 8  \\
\hline
$\tau_{\eta}$   & 0.211 & 0.215  & 0.213  & 0.206\\
$T_L^{xy}$  & 0.68 & 0.98  & 1.21 & 1.16 \\
\hline \hline
\end{tabular}
\caption{Kolmogorov time scale and lateral Lagrangian integral times. Data are for $Ra=10^7$ and
different aspect ratios. \label{kolscale}} 
\end{center}
\end{table}

Recent numerical Lagrangian studies in homogeneous isotropic turbulence by Sawford {\it et al.} 
\cite{Sawford2008} yielded a sensitive dependence of the pair dispersion on the initial pair separation. This was 
confirmed in the case of convective turbulence in an extended layer \cite{Schumacher2009}. In addition it was 
found there that the intermediate evolution of the dispersion in thermal convection depends on the seeding height 
of the tracer pairs since convective turbulence is inhomogeneous in the vertical direction. Here, we repeat
these studies. In Fig.~\ref{fig:tracer_R}(c), the pair dispersion normalized by 
$\langle\epsilon\rangle t^3$ is plotted for three different initial separations, namely $\eta$, $2\eta$ and $4\eta$, with 
$\eta=\nu^{3/4}/\langle\epsilon\rangle^{1/4}$ the Kolmogorov length. The figure shows that a plateau, i.e. $R^2\sim t^3$,  
is reached for the smallest initial separation only. 

The autocorrelation functions of the lateral and vertical velocities are shown in Fig.~\ref{fig:tracer_R}(d) for one run. 
The results  are in agreement with \cite{Schumacher2008}. The vertical velocity is strongly anticorrelated to the lateral one. This strong anticorrelation is in line with the rapid upward and downward motions as indicated 
by the signals of velocity components along a tracer path in Fig.~\ref{fig:tracer_t}, in which fluctuations in 
$u_z(t)$ are stronger than those in $u_x(t)$ and $u_y(t)$ (time signals of $u_x$ and $u_y$ are not shown here). In order to demonstrate this more quantitatively, we
plot in Fig.~\ref{fig:tracer_corr} the vertical displacement of the Lagrangian particles and the corresponding 
integrand of definition $T_L^{(z)}$ (see Eq.~\ref{eq:Tla})) for 6 tracers that are initially seeded close to bottom boundary layer. The 
pronounced minimum of $u_z(0)u_z(\tau)$ coincides with the reversal point of $z(t)$ close to the top boundary layer. 

Fig.~\ref{fig:tracer_R_Gamma} shows the dependence of the long-time behavior of the lateral dispersion
on the aspect ratio. Only for the largest aspect ratio ($\Gamma=12$) a diffusion limit with $R^2_{xy}(t)\sim t$ is obtained. It underlines an
important difference in contrast to configurations with periodic boundary conditions in the lateral directions. In our case, the dispersion is additionally constrained by the sidewalls for small-aspect-ratio systems.
 
\begin{figure}
\begin{center}
\includegraphics[angle=0,scale=0.48,draft=false]{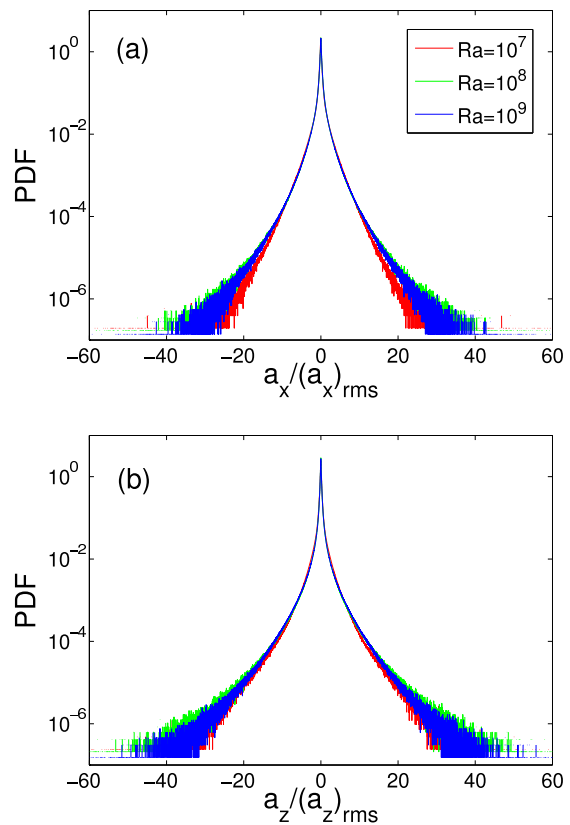}
\caption{(Color online) The probability density functions of the acceleration 
components for different Rayleigh numbers at $\Gamma=1$. Amplitudes 
are normalized by the corresponding root-mean-square value.  (a) 
Component $a_x$. (b) Component $a_z$.} 
\label{fig:pdf_ak}
\end{center}
\end{figure}
\begin{figure}
\begin{center}
\includegraphics[angle=0,scale=0.48,draft=false]{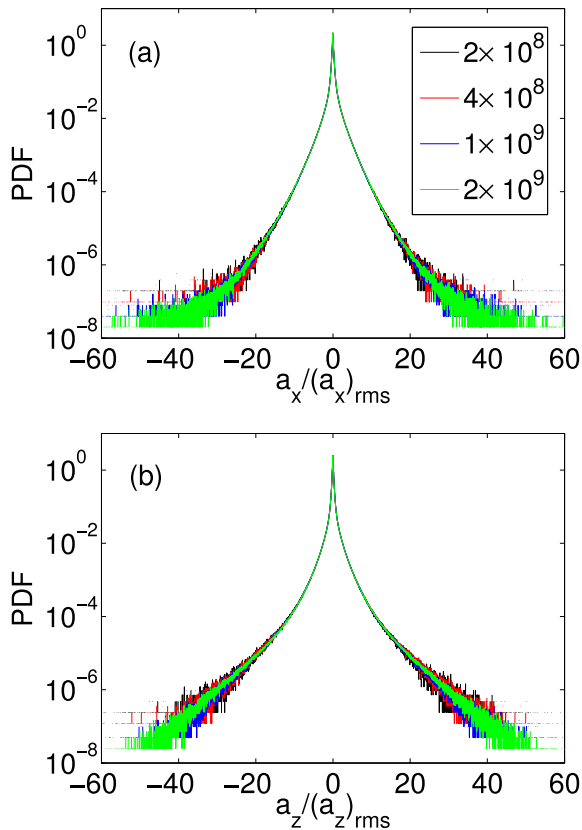}
\caption{(Color online) The convergence of the probability density functions 
of one lateral and the vertical acceleration components for different 
$Ra=10^7$ and $\Gamma=1$. Data are for $2\times 10^8$, $4\times 10^8$, 
$10^9$, and $2\times 10^9$ sample events.} 
\label{fig:pdfconv}
\end{center}
\end{figure}
\begin{figure}
\begin{center}
\includegraphics[angle=0,scale=0.48,draft=false]{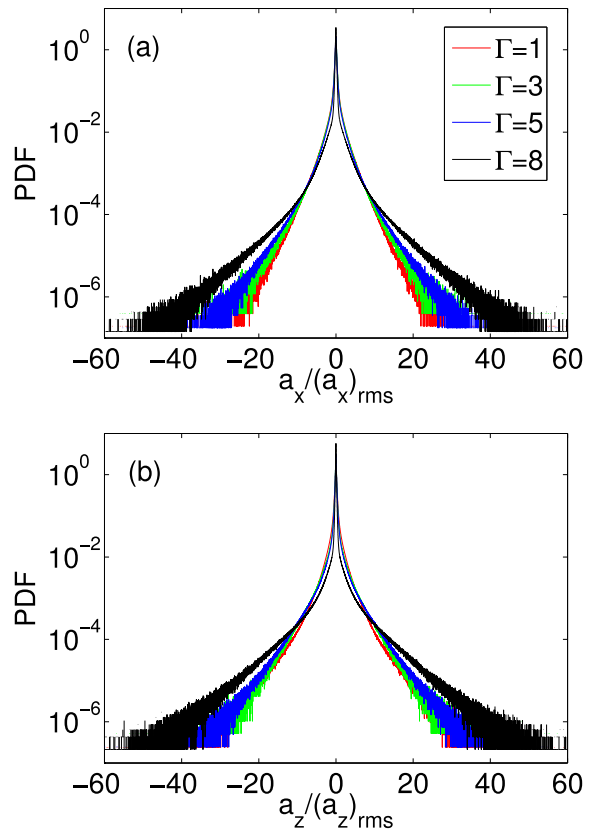}
\caption{(Color online) The probability density functions of the acceleration 
components. (a) Component $a_x$. (b) Component $a_z$. The uppermost 
graph is for the largest aspect ratio in both panels.}
\label{fig:pdf_ax_az_gamma}
\end{center}
\end{figure}

\section{Acceleration statistics}
Finally, we report the probability density function (PDF) of acceleration components along 
three-dimensional trajectories.  In Fig.~\ref{fig:pdf_ak}  we compare data for three different 
Rayleigh numbers with fixed aspect ratio $\Gamma=1$. Both lateral and vertical acceleration  components
show no systematic trend with $Ra$. The stretched exponential form is similar to those reported 
in laboratory  experiments \cite{Lap1,Mo1,Vo1} and DNS simulations of turbulent convection 
\cite{Schumacher2008,Schumacher2009}.  

A critical issue of numerical and experimental Lagrangian studies -- not only in turbulent 
convection -- is the statistical convergence, which requires to determine, for example, a 
fourth-order moment, such as the flatness $F(a_k)$. In Lagrangian turbulence, the situation 
becomes particularly problematic since the tails of the 
PDFs are found to be strongly scattered \cite{Sreenivasan2010,Mordant2004}. 
The statistical uncertainty is also manifested in the scattered tails 
of the PDFs in Fig.~\ref{fig:pdf_ak} although  more than $2\times 10^8$ data points have 
been included. In 
Fig.~\ref{fig:pdfconv}, we ran the simulation for significantly longer period such that the number of samples 
became one order of magnitude larger. The tails become less scattered and more extended 
with increasing number of samples. The data indicate that the convergence in the vertical direction
is slower than in the lateral direction. Nevertheless, the scatter for the fourth order moments 
remains so strong that the flatness is obtained with a big error bar only. The flatness is defined 
as
\begin{equation}
F(a_k)=\frac{\langle a_k^4\rangle_{L,t}}{\langle a_k^2\rangle_{L,t}}
\end{equation}
with $k=x,y$ and $z$. We found that $F(a_z)= 89\pm 18$ for $4\times 10^9$ samples.   

The dependence on the aspect ratio $\Gamma$ is demonstrated in Fig.~\ref{fig:pdf_ax_az_gamma}. It can be seen that the tails of the PDF of $a_x$ grow with increasing aspect ratio, which implies that less constrained 
lateral motion causes larger acceleration magnitudes. Meandering of the tracers among
different LSC rolls \cite{Bailon2009} will probably contribute to larger amplitudes in the 
acceleration as well. Interestingly, a similar but weaker trend holds for the vertical acceleration 
component as well.   

\section{Summary and discussion}
We have presented DNS studies on the Lagrangian tracer particle dynamics in turbulent 
Rayleigh-B\'{e}nard convection. The main objective of this work was to investigate how the 
finiteness of the cell size in convection  impacts the motion, dispersion and acceleration of
the tracer particles. We developed an interpolation scheme for the staggered computational
mesh, which was used in the Eulerian simulation. Our interpolation scheme reproduces accurately the global heat transfer in the Lagrangian frame $Nu_L$, which matches the corresponding value in the Eulerian frame $Nu_E$. We detected however large fluctuations of $Nu_L(t)$ about the mean value since the majority of the tracers is moving along the mean circulation path in the cell. Our studies suggest that either a large 
number of particles or a very long time integration period are necessary for faster convergence of $Nu_L(t)$. The convergence is found to be slower for a non-uniform initial tracer seeding.

In the pair dispersion analysis, we confirmed the initial ballistic regime 
($R^2\sim t^2$) and a very short-range regime
with $R^2\sim t^3$  which have also been found in other studies on convection \cite{Schumacher2008,Schumacher2009}
and simulations of homogeneous isotropic turbulence in a periodic cube \cite{Boffetta2002,Sawford2008}. The finite 
size of the convection cell suppresses the dispersion limit; hence a Taylor-like regime ($R^2\sim t$) is absent  for
$\Gamma<8$ in our studies as opposed to \cite{Schumacher2008}. The autocorrelations of the vertical velocity
component are strongly anti-correlated. This is in line with \cite{Schumacher2008} and seems to be
associated with the characteristic structures -- the thermal plumes -- present in convection. These structures have a strong influence on the vertical tracer motion. In particular they cause the reversals in the tracer tracks, namely  when a particle hits the top or bottom plates. While we did not 
observe a Rayleigh-number-dependence of the acceleration statistics, a systematic dependence of the PDFs of the acceleration components  on $\Gamma$ was obtained. To conclude,  the basic Lagrangian properties we studied in the present work agree qualitatively with those in previous studies, e.g. in \cite{Schumacher2008,Schumacher2009}. 
The additional variation of the aspect ratio is in line with changes in the large-scale circulation in the convection cell, which affects the Lagrangian 
tracer dispersion and acceleration. 

Our studies complemented the only existing Lagrangian laboratory experiment on convection by Gasteuil {\it et al.} \cite{Gasteuil2007}. 
We could confirm qualitatively some results obtained in their experiments. Further experimental
and numerical studies in the 
large aspect ratio regime are desirable and necessary. Similar to the Eulerian case, we can expect the aspect ratio dependence to disappear for sufficiently large values of
$\Gamma$.

\section*{Acknowledgements} 
We wish to thank Oleg Zikanov for discussions
and the J\"ulich Supercomputing Centre (Germany) for support with computing resources 
on the JUROPA cluster under grant HIL03. This work is also supported by the Deutsche 
Forschungsgemeinschaft (DFG) under grant SCHU1410/2-1 and by the Heisenberg Program 
of the DFG under grant SCHU 1410/5-1. JS acknowledges partial travel support by the European
COST Action MP0806 ``Particles in turbulence".

\end{document}